\documentclass[copyright,creativecommons]{eptcs}
\providecommand{\event}{QPL 2013}
\usepackage{breakurl}  
\usepackage{amsmath}
\usepackage{amsthm}
\usepackage{amsfonts}
\usepackage{amssymb}
\usepackage{enumerate}
\usepackage{mathrsfs}
\usepackage{stmaryrd}
\usepackage{paralist}
\usepackage{bm}
\input{diagxy.tex}

\renewcommand{\.}{\,.}

\newcommand{\cB}{\ensuremath{{\mathcal{B}}}}
\newcommand{\cC}{\ensuremath{{\mathcal{C}}}}

\newcommand{\cD}{\ensuremath{{\mathcal{D}}}}

\newcommand{\cF}{\ensuremath{{\mathcal{F}}}}

\newcommand{\cJ}{\ensuremath{{\mathcal{J}}}}

\newcommand{\cL}{\ensuremath{{\mathcal{L}}}}
\newcommand{\cM}{\ensuremath{{\mathcal{M}}}}

\newcommand{\cP}{\ensuremath{{\mathcal{P}}}}

\newcommand{\pc}[1]{\textit{#1}}
\newcommand{\pt}[1]{\texttt{#1}}
\newcommand{\mt}[1]{\mathtt{#1}}
\newcommand{\mi}[1]{\mathit{#1}}

\newcommand{\bbR}{\ensuremath{{\mathbb {R}}}}

\newcommand{\ga}{{\bm{a}}}
\newcommand{\gb}{{\bm{b}}}
\newcommand{\gcp}{{\bm{c}_{2}}}

\newcommand{\gn}{{\bm{n}}}
\newcommand{\gnp}{{\bm{n}_{2}}}

\newcommand{\gp}{{\bm{p}}} 
 
\newcommand{\gs}{{\bm{s}}}

\newcommand{\la}{\langle}
\newcommand{\ra}{\rangle}
\newcommand{\coname}[1]{\llcorner #1\lrcorner}
\newcommand{\name}[1]{\ulcorner #1\urcorner}

\newcommand{\lra}{\longrightarrow }

\newcommand{\oa}{\ensuremath{\overrightarrow}} 
\newcommand{\ol}[1]{\overline{\mathtt{#1}}}
 
\newcommand{\ot}{\otimes}

\newcommand{\s}[1]{\ensuremath{\left\{ #1 \right\}}}
\newcommand{\st}[2][n]{\ensuremath{#2_1  \dots  #2_#1}}
\newcommand{\se}[2][n]{\ensuremath{#2_1, \dots ,#2_{#1}}}
\newcommand{\sn}[2][n]{\ \text{for} \  #2= 1, \dots , #1}

 \newcommand{\bra}[1]{\langle  #1\vert }
\newcommand{\ba}{\begin{array}}
\newcommand{\ea}{\end{array}}
\newcommand{\bc}{\begin{cases}}
\newcommand{\ec}{\end{cases}}
\newcommand{\bco}{\begin{compactenum}[ ]}
\newcommand{\eco}{\end{compactenum}}
\newcommand{\bcp}{\begin{compactenum}[1)]}
\newcommand{\ecp}{\end{compactenum}}
\newcommand{\be}{\begin{equation}}
\newcommand{\ee}{\end{equation}}
\newcommand{\bi}{\begin{itemize}}
\newcommand{\ei}{\end{itemize}}
\newcommand{\pro}[1]{\begin{proof}#1\end{proof}}
\newcommand{\bn}{\begin{enumerate}}
\newcommand{\en}{\end{enumerate}}
\newcommand{\bes}{\begin{equation*}}
\newcommand{\ees}{\end{equation*}}
\newcommand{\re}[1]{(\ref{#1})}
\newcommand{\keywords}[1]{\textit{Keywords}:\begin{footnotesize}#1\end{footnotesize}}

\title{From Logical to Distributional Models}
\author{Anne Preller
\institute{LIRMM,\\Montpellier, France}
\institute{
\thanks{The author wishes to thank the LIRMM and in particular the research group TEXTE for support}\\
}
\email{preller@lirmm.fr}
}
\def\titlerunning{From Logical to Distributional Models}
\def\authorrunning{Anne Preller}

\begin{document}
\newtheorem{theo}{Theorem}
\newtheorem{prop}{Proposition}
\newtheorem{lem}{Lemma}
\newtheorem{coro}{Corollary}
\newtheorem{exa}{Example}
\newtheorem{defi}{Definition}
\newtheorem{rem}{Remark}
\newtheorem{fact}{Fact}
\maketitle

\begin{abstract}
The paper relates two variants of semantic models for natural language, logical functional models and compositional distributional vector space models, by transferring the logic and reasoning from the logical to the distributional models.
The geometrical operations of quantum logic are reformulated as algebraic operations on vectors.  A map from functional models to vector space models makes it possible to compare the meaning of sentences word by word.  
\end{abstract}
{\keywords  { compositional semantics for natural language, compact closed categories, quantum logic,  logical models, vector space logic, algebraic connectives}}
\section{Introduction}
Semantic models for natural language vary from logical models, e.g. first order models or Montague models, to conceptual models. Conceptual models englobe variants of higher order type theory \cite{as11} and the geometrical vector space models based on quantum logic   \cite{wi04},   \cite{wi08},   \cite{ri}. They all involve reasoning, an essential ingredient of compositional semantics, \cite{kr1}.

This paper traces the switch from logical functional models of pregroup grammars to the distributional vector space models of \cite{ccs} and shows how the predicate logic of functional models changes to the quantum logic of vector space models.  It also proposes a way to fill a gap left in \cite{ks}, namely logic.  

Pregroup grammars \cite{la99} provide the common mathematical background of compact closed categories, which facilitates the passage from the functional logical models to vector space models. Both kinds of models are implemented by structure preserving functors defined on the \emph{lexical category} $\cL(\cB)$, the free compact closed category generated by a partially ordered set of basic types $\cB$ and the entries in a pregroup dictionary $\cD_\cB$.   In this study, all functors map  entries in the dictionary to vectors in the category ${\bf FVect}_\bbR$ of finite dimensional vector spaces over the real numbers. 

Structure preserving functors are necessarily compositional, because all strings recognised by a pregroup grammar are morphisms of the lexical category. Indeed, any grammatical analysis produced by a pregroup grammar corresponds to a morphism $r$ of the lexical category.  The meaning of the string is defined in the lexical category as the composite of $r$ with the juxtaposition (tensor product) of the lexical entries (vectors). 
A model implemented by a functor from $\cL(\cB)$ to some compact closed category commutes with the structural operations. Hence the value of the string is computed from the values of the words using the  operations of the compact closed category.

The main difference between the functors standing for logical functional models and those standing for vector space models lies in the interpretation of the sentence type.  The former models map it to a two-dimensional space of truth-values and the latter to a higher dimensional space of concepts represented by words.

A vector space model interprets  the so-called  \emph{property words} or \emph{concepts}, i.e. nouns, adjectives and verbs, by vectors in a fixed finite dimensional space $W$, the basis vectors of which are identified with a set of basic concepts.  Words with a logical content like determiners, relative pronouns, connectives (\emph{and}, \emph{not} etc.) and auxiliary verbs are not interpreted. They are `{noise}'.  The compositional extension to meanings of strings is given in \cite{ks} ``via a strongly monoidal functor from the free pregroup \footnote{The authors seem to use the term `free pregroup' in a sense that is not compatible with the existence of a strongly monoidal functor, see Section \ref{VSM}.}  of basic types to the category of finite dimensional vector spaces ${\bf FVect}_\bbR$" using the Frobenius multiplication in form of the pointwise product $\odot$ for composition. 

Logical functional models interpret all words in the lexicon.  They map nouns to vectors with coefficients equal to $0$ or $1$, attributive adjectives to projectors  (i.e. linear maps that map any basis vector to itself or to $0$)  and verbs and predicative adjectives to predicates, (i.e. linear maps that take their values in a space of \emph{truth values}).   They interpret logical words as distinguished linear maps that operate on predicates by composition. For instance, if  $\pt{not}:S \rightarrow S$ is the linear map  interpreting the negation \emph{not} and \pt{blue}  the predicate corresponding to the adjective \emph{blue} then $\pt{not}\circ \pt{blue}$ is the meaning of \emph{not blue}. The  induced logic is  four-valued and strictly extends first order predicate logic.  Besides `true' and `false' it has truth-values `meaningless', e.g. \emph{rocks sleep}, and `mixed', e.g. \emph{rocks are grey}. 

The question addressed here is if and how the compositional approach to vector space models can be extended to noise words.  Indeed, the compositional approach via a functor does not extend to most logic words.  For instance, there is no vector $\oa{\mi{not}}$ for which $ \oa{\mi{not}}\odot \oa{\mi{blue}}$ would be orthogonal  to  $\oa{\mi{blue}}$.  

The geometrical logical connectives  \emph{and}, \emph{not} \emph{or}, \emph{if-then} of quantum logic operate on projectors, not on vectors.  
After reformulating the geometrical operations as algebraic operations, this study shows that the algebraic operations and the corresponding consequence relation define a conditional logic on the concept space similar to the conditional logic of information retrieval  in   \cite{ri}.  Negation, for instance, coincides with orthogonality. 

How adequate is the extension of the vector space model with respect to the meaning of sentences? To test the adequateness, I pair any logical functional model $\cF$ with a vector space model $\cM_C$ interpreting words in a space $C$ and define a map $\cJ$ from `properties', e.g. vectors, projectors and  predicates, to vectors of $C$ such that $\cM_C$ coincides on words with the composite  $\cJ\circ\cF$. The model $\cM_C$ depends on the model $\cF$, a possible world to which the statements refer, and the choice of basic vectors of $C$, expressing semantic relations between words,  and the grammatical role of words in strings of words. The coefficients of the vectors in $\cM_C$ have a truth probabilistic content related to and motivated by quantum logic.

The logical functional model and the vector space model are both functors defined on the lexical category.   The value in the functional model is the composite of the word values, e.g. $\cF(\emph{rocks are grey}) =\cF(\mt{are})\circ \cF(\mt{grey}) \circ \cF(\mt{rocks}) $, whereas the vector space model $\cM_C$ uses the pointwise multiplication, e.g. $\cM_C(\emph{rocks are grey})$ = $\cM_C(\mt{rocks})\odot \cM_C(\mt{are})\odot\cM_C(\mt{grey})$. 
The meanings of strings in the two models can be compared word by word and operator by operator.  Under sufficient conditions, the truth-probabilistic content of words is preserved by strings, i.e.
$\cJ(\cF(\mi{word}_1 \dots \mi{word}_n)) = \cM(\mi{word}_1 \dots \mi{word}_n)$.  

In fact, the map $\cJ$ preserves truth and reflects the algebraic consequence relation.  Moreover, $\cJ$ preserves negation. 
Under sufficient conditions $\cJ$ also preserves the binary connectives. In particular, under these conditions the pointwise product commutes with the probability interpretation, because the conjunction of vectors is their pointwise product. The algebraic operations also coincide with the connectives of predicate logic in the degenerate case where individuals play the role of `basic concepts'. 
\section{Lexical semantics of pregroup grammars}\label{SYNSEM}
\renewcommand{\to}{\rightarrow}
The free pregroup $\cP(\cB)$ generated by a partially ordered set $\cB$ introduced in   \cite{la99} is a partially preordered monoid in which every element $a$ has a right adjoint $a^r$ and a left adjoint $a^\ell$ characterised by the equalities
$$ a a^r \leq 1\leq a^r  a \qquad a^\ell a\leq 1\leq a a^\ell\.$$
 Semantics requires a slightly modified definition, namely the free compact closed category $\cC(\cB)$ generated by  $\cB$, introduced in \cite{pl}. The only difference between the two versions is that  $\cP(\cB)$  identifies all morphisms of $\cC(\cB)$ that have a common domain and a common codomain.  This difference is essential when compositional semantics in vector spaces are mediated by a functor. An example of  morphisms identified in $\cP(\cB)$ leading to different meanings is given in \cite{pl}.  See also Fact \ref{nofunctor} in Section \ref{VSM} showing that a strongly monoidal functor from the free pregroup maps every element of the free pregroup to a space of dimension at most one. 

Recall that a \pc{monoidal category} consists of a category $\cC$, a bifunctor,  (denoted somewhat misleadingly by the tensor symbol) $\ot$, a distinguished object $I$, the \emph{unit  of the bifunctor}, and natural isomorphisms $\alpha_{ABC}:(A \ot B) \ot C \to A \ot (B \ot C)$,  $\lambda_A : A \to I \ot A$ and $\rho_A : A \to  A \ot I$ subject to the coherence conditions of   \cite{mcl}. A monoidal category is \pc{symmetric} if there is a natural isomorphism $\sigma_{AB}: A \ot B \to B \ot A$ such that $\sigma_{AB}^{-1}=\sigma_{BA}$,  again subject to the coherence conditions of   \cite{mcl}. 
A monoidal category is \pc{compact closed} if for every object $A$ there are objects  $A^r$ and $A^\ell$, called \emph{right adjoint} and \emph{left adjoint} of $A$ respectively, and morphisms $\eta_A:I\to A^r \ot A$, $\epsilon_A:A\ot A^r \to I$, $\eta_{A^\ell}:I\to A \ot A^\ell$, $\epsilon_{A^\ell}:A^\ell\ot A  \to I$ satisfying 
\bes\label{adjoint}\lambda_A^{-1}\circ (\epsilon_A \ot 1_A)\circ \alpha_{A\,A^r\,A}^{-1} \circ (1_A \ot \eta_A) \circ \rho_A  = 1_A,\quad
 \lambda_{A^\ell}^{-1} \circ   (\epsilon_{A ^\ell} \ot 1_{A ^\ell})\circ \alpha_{A^\ell\,A \,A^\ell}^{-1} \circ (1_{A ^\ell} \ot \eta_{A ^\ell}) \circ \rho_{A^\ell} = 1_{A^\ell}.\ees
For notational convenience, the associativity isomorphisms $\alpha_{ABC}$ and the unit isomorphisms $\lambda_A$ and $\rho_A$ are replaced by identities, e.g.  $(A \ot B) \ot A = A \ot (B \ot C)$, $A = I \ot A$ and $ A \ot I$\. The equalities above become the \emph{ adjoint  equalities}
\be\label{adjoint}(\epsilon_A \ot 1_A) \circ (1_A \ot \eta_A)   = 1_A,\quad
(\epsilon_{A ^\ell} \ot 1_{A ^\ell}) \circ (1_{A ^\ell} \ot \eta_{A ^\ell})  = 1_{A^\ell}.\ee

 A functor between compact closed categories  \emph{preserves the compact closed structure} if it commutes with the tensor product, the tensor unit and right and left adjoints up to natural isomorphisms. 

Every morphism of the free compact closed category generated by some  category $\cC$ can be designated by a `normal' graph where all links are labelled by morphisms  of $\cC$ and all paths have length $1$. In the case where the label is an identity it may be omitted. The graph displays the domain above, the codomain below. Vertical links correspond to right or left adjoints of morphisms of $C$,  overlinks to names and the underlinks to conames of morphisms of $C$.   For example,
the right and the left adjoint of a basic morphism $f:\ga \to \gb$ are represented by the graphs
$$f^r:\gb^r \to \ga^r= 
\bfig
\place(0,0)[\gb^r]\place(0,-330)[\ga^r]
\morphism(-20, -40)|r|/<-/<0,-230>[` ;f]\efig
\qquad \qquad 
f^\ell:\gb^\ell \to \ga^\ell= 
\bfig
\place(0,0)[\gb^\ell]\place(0,-330)[\ga^\ell]
\morphism(-20, -40)|l|/<-/<0,-230>[` ;f]\efig
$$
 Any morphism $f:\ga \to \gb$  has two names
$$\ba{ll}\text{right name}\\
\ba{ll}
\eta_f  \quad &=(f ^r\ot 1_{\gb} ) \circ \eta_{\gb }\\
\bfig
\place( 0,450)[I]
\morphism(-130, 50)|a|/{@{->}@/^.6em/}/<270, 0>[  `; f ]
\place(0,0)[\ga^r\ \ot \  \ \gb  ]
\efig&
=\bfig\place(0,250)[I]
\morphism(-140, -40)/<-/<0,-240>[` ;f]
\morphism(140, -40)/->/<0,-240>[` ; ]
\morphism(-130, 60)|a|/{@{->}@/^.6em/}/<270, -10>[  `; ]
\place(0,0)[ \gb^r\ \ot \  \ \gb ]
\place(0,-340)[\ga^r\ \ot  \ \ \gb  ]
\efig\ea 
\ea
\quad 
\ba{ll}
\text{left name}\\
\ba{ll}
\eta_{f^\ell}&=(f \ot  1_{\ga^\ell}) \circ \eta_{\ga^\ell}\\
\bfig\place( 0,450)[I]
\morphism(-150, 60)|a|/{@{<-}@/^.6em/}/<270, -10>[  `;f ]
\place(0,0)[ \gb\ \ \ot \  \ga^\ell ]\efig&=\bfig\place(0,250)[I]
\morphism(-150, -40)|a|<0,-230>[` ;f]
\morphism(130, -40)/<-/<0,-240>[` ; ]
\morphism(-150, 60)|a|/{@{<-}@/^.6em/}/<270, -10>[  `; ]
\place(0,0)[ \ga\ \ \ot \ \   \ga^\ell ]
\place(0,-340)[\gb\ \ \ot \ \   \ga^\ell ]
\efig\ea
 \ea
\.$$
Analogous definitions and notations apply to conames, namely $\epsilon_f$ and $\epsilon_{f^\ell}$
$$\ba{ll}\text{right coname}\\
\ba{cc}
\epsilon_f  \quad &=\epsilon_{\gb }\circ (f \ot 1_{\gb^r} ) \\
\bfig
\place( 0,-450)[I]
\morphism(-130, -50)|b|/{@{->}@/_.6em/}/<270, 20>[  `; f ]
\place(0,0)[\ga^r\ \ot \  \ \gb  ]
\efig&
=\bfig
\place(0,-550)[I]
\morphism(-140,  -40)/->/<0,-230>[` ;f]
\morphism(100,  -40)/<-/<0,-230>[` ; ]
\morphism(-130, -370)|a|/{@{->}@/_.6em/}/<270, 20>[  `; ]
\place(0,0)[ \ga\ \ot \  \ \gb^r ]
\place(0,-330)[\gb\ \ot  \ \ \gb^r  ]
\efig\ea 
\ea
\quad 
\ba{ll}
\text{left coname}\\
\ba{ll}
\epsilon_{f^\ell}&=  \epsilon_{\gb^\ell}\circ(  1_{\gb^\ell}\ot f)\\
\bfig\place( 0,-450)[I]
\morphism(-130, -40)|b|/{@{<-}@/_.6em/}/<270, 20>[  `;f ]
\place(0,0)[ \gb^\ell\  \ot \ \  \ga ]\efig
&=\bfig\place(0,-550)[I]
\morphism(-140,  -40)/<-/<0,-230>[` ;]
\morphism(130,  -40)/->/<0,-230>[` ;f ]
\morphism(-130, -370)|a|/{@{<-}@/_.6em/}/<270, 20>[  `; ]
\place(0,0)[ \gb^\ell\ \ot \  \ \ga ]
\place(0,-330)[\gb^\ell\ \ot  \ \ \gb   ]
\efig\ea
 \ea
\.$$

The equality of the composite graph on the right to the normal graph on the left  is a particular instance of the so-called `yanking'. Start at the tail of any link situated in the top or bottom line and follow the oriented links until a head situated in the top or bottom line is reached. Replace the whole path by a single link, labelled by the composite of the labels in the order they  are encountered. In the case of names, yanking works by definition. The general  case concerning the composite of two arbitrary graphs follows from the adjoint axioms \re{adjoint}.
$$(\epsilon_\ga \ot 1_\ga) \circ (1_\ga \ot \eta_\ga)   =  
\bfig
\place(0,0)[ \ga \ \ \ot\ \ \ga^r \ \ \ot\ \ \ga]
\place(0,320)[\ga]
\place(0,-320)[\ga]
\morphism(0,300)|m| /->/<-250, -260>[  `;]
\morphism(260,-20)|m| /->/<-250, -260>[  `;]
\morphism(-260, -40)|m|/{@{->}@/_.6em/}/< 260,20>[  `;]
\morphism(0, 40)|m|/{@{->}@/^.6em/}/< 260, 0>[  `;]
\efig=\bfig
\place(0,320)[ \ga]  
\place(0,-320)[\ga]
\morphism(0,280)|m| /->/<0, -560>[  `;]
 \efig
=  
\bfig
\place(0,0)[ \ga\ \ \ot\ \ \ga^\ell \ \ \ot\ \  \ga]
\place(0,320)[\ga]
\place(0,-320)[\ga]
\morphism(0,300)|m| /->/<250, -260>[  `;]
\morphism(-260,-20)|m| /->/<250, -260>[  `;]
\morphism(-20, -40)|m|/{@{<-}@/_.6em/}/< 290,20>[  `;]
\morphism(-260, 30)|m|/{@{<-}@/^.6em/}/< 260, 0>[  `;]
 \efig=(\epsilon_{\ga ^\ell} \ot 1_{\ga ^\ell}) \circ (1_{\ga ^\ell} \ot \eta_{\ga ^\ell})$$

The same morphisms composed in the opposite order result in the normal graph on the right
 $$  ( 1_{\ga} \ot \eta_{\ga })\circ   (\epsilon_\ga \ot 1_{\ga} )=
\bfig
\place(0,450)[\ga \ \  \ot \ \ga^r\ \ot \ \  \ga  ]
\place(0,-450)[\ga \ \ \ot \ \ \ga^r\ \ \ot \  \ \ga ]
\place(0, 0)[\ga]
\morphism(300, 450)|m|<-280, -390>[ ` ; ]
\morphism(0, -410)|a|/{@{->}@/^.6em/}/<300, 0>[  `;   ]
\morphism(-290, 410)|b|/{@{->}@/_.6em/}/<310, 20>[  `;   ]
\morphism( 10, -10)|m|<-280, -380>[ ` ; ]
\efig= \bfig
\place(0,450)[\ga \ \ \ot\  \ \ga^r\ \ \ot\  \  \ga  ]
\place(0,0)[\ga \ \ \ot \ \  \ga^r\ \ \ot \  \ \ga ]
\morphism(300, 430)|m|<-570, -370>[ ` ; ]
\morphism(0, 40)|a|/{@{->}@/^.6em/}/<300, 0>[  `;   ]
\morphism(-290, 410)|b|/{@{->}@/_.6em/}/<310, 20>[  `;   ]
\efig\.
  $$ 
  
The free compact closed category generated by an arbitrary category has the `normal form property', namely there is a one-to-one correspondence between morphisms and graphs where all paths are reduced to a single link.  This implies that 
\bes\label{semanticdiff}1_{\ga\ot \ga^r \ot \ga}= 
\bfig
\place(0,450)[\ga \ \ \ot\ \ \ga^r\ \ \ot \  \ga  ]
\place(0,0)[\ga\ \ \ot\  \ \ga^r\ \ \ot \  \ga ]
\morphism(-280, 410)|l|<0, -350>[ `;  ]
\morphism(0, 50)|l|<0, 350>[ ` ; ]
\morphism(280, 410)|l|<0, -350>[ ` ; ]
\efig
\neq   
\bfig
\place(0,450)[\ga \ \  \ot \ \ \ga^r\ \ \ot \  \ \ga  ]
\place(0,0)[\ga \ \  \ot\ \ \ga^r\ \ \ot\  \  \ga ]
\morphism(300, 430)|m|<-590, -380>[ ` ; ]
\morphism(0, 40)|a|/{@{->}@/^.6em/}/<300, 0>[  `;   ]
\morphism(-290, 410)|b|/{@{->}@/_.6em/}/<310, 20>[  `;   ]
\efig= ( 1_{\ga} \ot \eta_{\ga })\circ   (\epsilon_\ga \ot 1_{\ga} )\.\ees
 
The objects of the free compact closed category $\cC(\cB)$ generated by a partially ordered set $\cB$ are called \pc{types}, among them are the elements of $\cB$, called  \pc{basic types}. A  \pc{simple} type has the form 
$\dots\,, \ga ^{ (-2)} =\ga^{\ell\ell },\, \ga ^{ (-1)} =\ga^{\ell },\, \ga ^{ (0)} =\ga,\,\ga ^{ (1)} =\ga^{r },\, \ga ^{ (2)} =\ga^{rr },\ \dots$ where the $\ga$ is a basic type.  Any object of $\cC(\cB)$ can be written as a finite string of simple types where juxtaposition plays the role of the monoidal bifunctor. Clearly, $\cC(\cB)$ is not symmetric

A pregroup lexicon consists of pairs $\mi{word}:T$ where $\mi{word}$ is a word of natural language and $T$ a type. One can view each entry $\mi{word}:T$  as a formal expression $\ol{word}:I \to T$ in the language of compact closed categories.  For example, 
\bes\label{lex}
\ba{l@{\ : \  }l}
\mathit{no}  &  \gs\ot\gs^\ell \ot\gnp\ot\gcp^\ell      \\
\mathit{new}  & \gnp\ot\gcp^\ell  \\
\mathit{triangles}  & \gcp     \\
\mathit{are}  &  \gnp^r\ot\gs \ot\gp^\ell\ot\gnp    \\
\mathit{blue} &  \gn^r\ot  \gp
\ea
\quad 
\ba{l@{\ :I\to  }l}
\ol{no}&  \gs\ot\gs^\ell \ot\gnp\ot\gcp^\ell      \\
\ol{new}& \gcp\ot\gcp^\ell  \\
\ol{triangles}& \gcp     \\
\ol{are}&  \gnp^r\ot\gs \ot\gp^\ell\ot\gnp    \\
\overline{\pt{blue}}&  \gn^r\ot  \gp\ \.
\ea\ees
Here $ \gcp \leq\gnp \leq\gn$ are the basic types standing for plural common nouns, plural noun phrases and noun phrases where the number does not matter, in that order.
The basic types $\gp \leq \gs$  correspond to predicative adjectives and to sentences. The reader can find a more comprehensive grammar of English in   \cite{la08}.

Every lexical entry $\ol{word}:I \to T=\ga_1^{(z_1)}\ot\dots\ot\ga_n^{(z_n)}$ creates a `lexical  morphism' $\mt{word}:\ga_{i_1}\ot\dots\ot\ga_{i_m} \to \ga_j$ where  $z_{i_k}$ is odd for $k=1,\dots,m$ and  $z_j$ is even. For example, 
 $$
\ol{blue} = 
\bfig \place( 0, 340)[I]
 \place(0,0)[\gn^r \  \ot \, \  \gp]
 \morphism(-120,40)|m|/{@{ ->}@/^.6em/}/< 250, 0>[  `;]
 \place(-10, 140)[\scriptstyle{\pt{blue}}]
  \efig,\quad\bfig\morphism (380,50)|r| /<-/<0, 400>[ \gp `\gn;\pt{blue}]\efig \qquad \qquad 
 \ol{new} = 
\bfig \place( 0, 340)[I]
 \place(0,0)[\gcp \ \ot \,  \ \gcp^\ell]
 \morphism(-150,40)|m|/{@{<-}@/^.6em/}/< 260, -10>[  `;]
 \place(-10, 140)[\scriptstyle{\pt{new}}]
\efig,\quad\bfig\morphism (380,50)|r| /<-/<0, 400>[ \gcp `\gcp;\pt{new}]\efig 
 \qquad \qquad \ol{triangles} = \bfig \place(-10, 400)[I]\place(0,  0)[\gcp]\morphism ( -20,50)|r| /<-/<0, 300>[  `;\pt{triangles}]\efig$$
 $$
\ol{are}=
(1_{\gnp^{r}} \ot \eta_{\mt{are}^\ell}\ot 1_{\gn})\circ \eta_{\gnp} =
\bfig\place(0, 300)[I]
\place( 0,0)[  \gnp ^r\  \ot \ \gn ]
\morphism(- 130,40)|a|/{@{>}@/^.5em/}/<260,-10>[ `;]
\place( 0,-350)[  \gnp ^r\  \ot\gs\,\  \ot\gp^\ell \ \ot \gn ]
\morphism( -90,-320)|a|/{@{<-}@/^.5em/}/<230, 0>[ `;\scriptstyle{\pt{are}}]
\morphism(-150, -50)/<-/<-190, -270>[ `;]
\morphism(130, -10)/>/<190, -300>[ `;]
\efig=\quad
\bfig\place(0, 400)[I]
\place( 0,0)[  \gnp ^r\ \ot \gs\,\ \ot \, \ \gp^\ell \ \ot \ \gn  ]
\morphism(- 330,40)|a|/{@{>}@/^1.2em/}/<670, -10>[ `;]
\morphism( -100,30)|a|/{@{<-}@/^.5em/}/<230, 0>[ `;]
\place(10,120)[\scriptstyle{\mt{are}}]
\efig ,\qquad\bfig\morphism (380,50)|r| /<-/<0, 400>[ \gs `\gp;\pt{are}]\efig$$ 
\bes 
 \ol{no} =\eta_{\mt{not}^\ell}\ot \eta_{\gnp^\ell}= 
 \bfig
 \place(0,250)[I]\place(0,0)[\gs \ \ \ot \  \  \gs^\ell\ \ \ot \ \ \gnp\ \  \ot  \ \, \gnp^\ell ]
  \morphism(-440,40)|m|/{@{<-}@/^.6em/}/< 230, 0>[  `;] 
  \place(-300, 150)[\scriptstyle{\pt{not}}]
 \morphism(90,40)|m|/{@{<-}@/^.6em/}/< 270, 0>[  `;]\efig,\qquad\bfig\morphism (380,50)|r| /<-/<0, 400>[ \gs `\gs;\pt{no}]\efig \.
\ees
Labelled links correspond to lexical morphisms like $\pt{not}:\gs \to \gs$, $\pt{blue}:\gn \to \gp$ and $\pt{new}:\gcp \to \gnp$.  Unlabelled links correspond to  (in)equalities of basic types $\mi{in}_{\ga\gb}:\ga \to \gb$.

Thus, every pregroup lexicon determines a \emph{lexical category} $\cL(\cB)$, namely the free compact closed category generated by the partially ordered set of basic types and the lexical morphisms.   
 A \pc{basic morphism} is an   (in)equality $\pc{in}_{\ga\gb}:\ga \to \gb$ between elements of $\cB$ or a lexical morphism $\mt{word}:\ga_1\ot\dots\ot\ga_m \to \gb$.  Note that the set of basic morphisms does not form a category.  The generating category of $\cL(\cB)$ is the monoidal category generated by $\cB$ and the lexical morphisms.   
 The lexical category is the mathematical tool of a pregroup grammar that provides both grammatical analysis and meanings of words and strings of words.  

Call \emph{reduction} any morphism $r:T_1 \dots T_n\to \gb$ of the free compact closed category involving (in)equalities or counits of (in)equalities of basic types only.  We omit all labels in the graphical picture, because they are uniquely determined by the tail and the head of the link. The basic type at the tail is necessarily less or equal to the basic type at the head.  

A  string of words $\st{{word}}$ is \pc{grammatical} if there are entries $\mi{word}_i: T_i$  in the lexicon,  a basic type $\gb$ and a reduction  $r:T_1 \dots T_n\to \gb$. 
 For instance. 
\bes
r=\bfig
\place(-800,200)[\pc{No}]\place(-170,200)[\pc{triangles}]\place(450,200)[\pc{are}]\place(1200,200)[\pc{blue}]
\place(40,0)[\gs\,\ \ot\, \ \gs^\ell\ \ \ot\ \gnp\ \,\ot\  \gnp^\ell\ \ \ot\ \, \gcp \, \ \ot\, \ \gnp^r \ \ot\   \,\gs\ \,\ot\ \,\gp^\ell  \  \ot \, \ \gn \,\ \ot\,\ \gn^r \ \ot\, \ \gp]
\morphism(-1280,-40)|m| /->/<0, -320>[  `;]
\place(-1280,-400)[\gs]
\morphism(-1030, -40)|m|/{@{<-}@/_2.2em/}/< 1390,20>[  `;]
\morphism(-770, -50)|m|/{@{->}@/_1.4em/}/< 830,30>[  `;]
\morphism(-520, -40)|m|/{@{<-}@/_.6em/}/< 330,20>[  `;]
\morphism(580, -60)|m|/{@{<-}@/_1.4em/}/< 760,20>[  `;]
\morphism(840, -40)|m|/{@{->}@/_.6em/}/< 260,30>[  `;]
\efig\ees
 
Meanings of grammatical strings are best defined abstractly in the lexical category $\cL(\cB)$. The \pc{meaning} of the string $\st{{word}}$ recognised by the reduction $r$ is  
$$r   \circ (\ol{word}_1 \ot \dots  \ot \ol{word}_n)\,. $$
For example, 
\bes\label{congraph}
\ba{c}
r\circ(\ol{no}\ot \ol{triangles}\ot \ol{are}\ot \ol{blue})=\\
\\
{\bfig
\place(40,0)[\gs\,\ \ot\, \ \gs^\ell\ \ \ot\ \gnp\ \,\ot\  \gnp^\ell\ \ \ot\ \, \gcp \, \ \ot\, \ \gnp^r \ \ot\   \,\gs\ \,\ot\ \,\gp^\ell  \  \ot \, \ \gn \,\ \ot\,\ \gn^r \ \ot\, \ \gp]
\morphism(-1280,-40)|m| /->/<0, -320>[  `;]
\place(-1280,-400)[\gs]
\morphism(-1030, -40)|m|/{@{<-}@/_2.2em/}/< 1390,20>[  `;]
\morphism(-770, -50)|m|/{@{->}@/_1.4em/}/< 830,30>[  `;]
\morphism(-520, -40)|m|/{@{<-}@/_.6em/}/< 330,20>[  `;]
\morphism(580, -60)|m|/{@{<-}@/_1.4em/}/< 760,20>[  `;]
\morphism(840, -40)|m|/{@{->}@/_.6em/}/< 260,30>[  `;]
\morphism(-1280, 40)|m|/{@{<-}@/^.6em/}/< 260, 0>[  `;]
\place(-1130,140)[\mt{\scriptstyle{not}}]
\morphism(-770, 40)|m|/{@{<-}@/^.6em/}/< 240, 0>[  `;]
\morphism ( -210, 50)|m|/<-/<0, 300>[  `I ; \mt{triangles} ]
\morphism(70, 40)|m|/{@{->}@/^1.4em/}/< 750,-10>[  `;]
\morphism(350, 30)|m|/{@{<-}@/^.6em/}/< 210,0>[  `;]
\place(430,130)[\scriptstyle{\pt{are}}]
\morphism(1090, 40)|m|/{@{->}@/^.6em/}/< 260, 0>[  `;]
\place(1190,140)[\mt{\scriptstyle{blue}}]
\efig}\\
 = \mt{not}\circ \mt{are} \circ \mt{blue}\circ\mi{in}_{\gnp\gn} \circ\mi{in}_{\gcp\gnp} \circ \mt{triangles}\\
  = \mt{not}\circ \mt{are} \circ \mt{blue}  \circ\mi{in}_{\gcp\gn} \circ \mt{triangles}.
 \ea\ees

Any compact closed structure preserving functor from the lexical category  $\cF$ into an arbitrary compact closed category $\cC$ provides us with a compositional interpretation of grammatical strings. The value of the string is computed from the values of the words with the help of the operations (tensor, composition etc.) of the category. 
 Assume, for instance, that $\cF$ is a compact closed structure preserving functor satisfying
$$\ba{c}N=\cF(\gn)=\cF(\gnp)=\cF(\gcp),   \qquad S =\cF(\gs) =\cF(\gp),\\
\cF(\mi{in}_{\ga\gb}) =1_{\cF(\gb)},\qquad  \cF(\mt{are})=1_S\.\ea$$
Then 
\be\label{meaning}\cF(r\circ(\ol{no}\ot \ol{triangles}\ot \ol{are}\ot \ol{blue}))= \cF(\mt{not}) \circ \cF(\mt{blue}) \circ \cF(\mt{triangles})\.\ee
\section{Vector space models as functors}\label{VSM}
A functor between monoidal categories  is \emph{strongly monoidal} if it commutes with the tensor product and the tensor unit up to natural isomorphisms. If both categories are compact closed a strongly monoidal functor  also commutes with the units and the counits of adjunction up to natural isomorphisms, hence the functor  preserves the compact closed structure.

The vector space models of   \cite{ks} are  "strongly  monoidal  functors from the  free pregroup to the full subcategory ${\bf FVect}_W$ of ${\bf FVect}_\bbR$ formed by the tensor powers of some chosen space $W$".   The subcategory ${\bf FVect}_W$ is a compact closed subcategory of ${\bf FVect}_\bbR$.  If $W$ has dimension greater than one, however, such a functor does not exist. 

\begin{fact}\label{nofunctor}There is no monoidal structure preserving functor from  $\cP(\cB)$  to ${\bf FVect}_\bbR$ that maps a basic type to a space of dimension greater than one. 
\end{fact}
\pro{
Suppose to the contrary that $A=\cF(\ga)$ has dimension at least two and that $\cF$ is a functor from  $\cP(\cB)$ to ${\bf FVect}_\bbR$ that preserves the monoidal structure and, as a consequence, also the compact structure up to natural isomorphisms.   
Then from  $1_{\ga\ot \ga^r \ot \ga}=  ( 1_{\ga} \ot \eta_{\ga })\circ   (\epsilon_\ga \ot 1_{\ga} )$ in $\cP(\cB)$ follows that
 $f= ( 1_A \ot \eta_A )\circ   (\epsilon_A  \ot 1_A )  $ is an isomorphism in ${\bf FVect}_\bbR$.  Assume $a_1, a_2 \in A$ are different basis vectors.  Then $(\epsilon_A  \ot 1_A )(a_1 \ot a_2 \ot a_1) =0$, because the counit $\epsilon_A :A\ot A \to I$ is the inner product of the space $A$.  It follows that $f(a_1 \ot a_2 \ot a_1)=0$ and therefore has no inverse. 
}
This problem disappears when the free pregroup $\cP(\cB)$ is replaced by the free compact closed category  $\cC(\cB)$.  A functor defined on $\cC(\cB)$, however,  is not eough, because it does not interpret the lexical entries. Adjectives in attributive position, for instance, have type $\gcp \ot\gcp^\ell$.  The left unit is the only morphism of $\cC(\cB)$ with domain $I$ and codomain $\gcp\ot \gcp^\ell$, because the  identity is the only endomorphism of the basic type $\gcp$ in the free category.  We must add the lexical morphisms and define a functor on the lexical category $\cL(\cB)$.  

Choosing a vector space $V$ and defining the word vectors in $V$ from a corpus is a complex task.  Here, we are only interested in the result, namely a  map $M_V$ from the entries of a pregroup lexicon to vectors in $V$, called \emph{vector space  model}.  The compositional extension of a $M_V$ via a functor such that the meaning of a string is again a vector in $V$, namely the pointwise product of the word vectors has a straight forward definition, due to the following fact.
\begin{fact}
Let $V$ be a  commutative monoid with binary operation $\star $ and neutral element $e$.  Then the following definitions define a compact closed category $ V_\star$
 \bco
 \item $V$ is the unique object of $V_\star$  
  \item  The elements $v \in V$ are the morphisms $v:V \to V$ of $V_\star $ 
  \item  $v_1 \circ v_2 =v_1 \star   v_2$, $1_V=e$
  \item $V\ot V =V$  
 \item $v_1 \ot v_2 =v_1 \star   v_2$  
\item $V^r = V=V^\ell$, $\eta_V=e=\epsilon_V$
 \eco
With these definitions, every morphism of $V_\star$ is equal to its name and coname.
\end{fact} 
\pro{The proof is straight forward. To see that $V_\star$ is a monoidal category with tensor unit $V$ we must show $\star $ is a bifunctor from the product category  $V_\star\times V_\star$ into  $V_\star$, that is to say we must show the two equalities
$$1_V\ot 1_V= 1_{V \ot V}\qquad (v_1\ot v_2)\circ (w_1\ot w_2)= (v_1\circ w_1)\ot (v_2\circ w_2)\.$$
The first follows immediately  from the definition.  The second holds, because the operation $\star $ is commutative.
Compact closure is as easily checked. Indeed,
 $$\ba{c}\eta_V=1_V: V \to V=V\ot V, \ \epsilon_V=1_V: V =  V\ot V \to V  \\
(\epsilon_V \ot 1_{V} ) \circ   ( 1_{V} \ot \eta_{V })= (e \star  e ) \star  (e \star  e )=e =1_V\.\ea$$ 
The last assertion that $\name m=m=\coname m$ is a straight forward consequence of the definitions.
}
If $\star$ is a commutative binary operation on a vector space $V$ then any compact closed structure preserving functor from the lexical category to $V_\star$ is a vector-based model in the sense of   \cite{ml}.

There are two obvious choices for the binary operation $\star$ in the case of  a  finite dimensional vector space $V$. One is the familiar addition of  vectors. The other one is the pointwise product $\odot$ of vectors used in   \cite{gs11}.   It is defined in terms of any orthonormal basis $A=\{a_1, \dots, a_n\}$  of $V$ by the equality
$$(\sum _{i=1}^n \alpha_i a_i) \odot (\sum _{i=1}^n \beta_i a_i)=\sum _{i=1}^n (\alpha_i \beta_i)a_i\.$$ 
The neutral element of $\odot$ is the vector $\oa 1 =\sum _{i=1}^n a_i$. 
We shall see that the pointwise multiplication also equips the vector space $V$ with a logic that is the vector version of the geometrical logic of projectors. 
  
Any vector space  model $M_V$ extends to a unique compact closed structure preserving functor $\cM_V$ from the lexical category $\cL(\cB)$ into the compact closed  category  $ V_\odot$ satisfying 
$$\ba{c}
\cM(\ga) =V \qquad\cM(\mi{in}_{\ga\gb})=\oa 1, \ \text{for all} \ \ga,\gb \in\cB\\
\cM(\mt{word}_T)= M(\mi{word}:T)\in V \ \text{for all lexical entries} \ \mi{word}:T \.\ea $$
Indeed, such a functor exists and is unique because the lexical category is the free compact closed category generated by the lexical words and the inequalities of basic types. The subscript $V$ will be omitted unless this leeds to confusion.

Note that the functor $\cM$ satisfies $\cM(T)=V$ for every type $T$.  Moreover,  $\cM(r) = \oa 1$ for any reduction $r:T_1 \ot \dots\ot T_n \to \gb$ of the pregroup grammar, because  a reduction is an expression of monoidal categories involving only conames of inequalities of basic types. Every morphism of $V_\odot$ is equal to its name(s). Hence, the value assigned to a grammatical string $\mi{word}_1 \dots\mi{word}_n$ is simply the pointwise product of the vectors 
\be\label{pointwise}
\cM(r   \circ (\ol{word}_{1T_1} \ot \dots  \ot \ol{word}_{nT_n}))= M(\mi{word}_1:T_1)\odot \dots\odot  M(\mi{word}_n:T_n)\.
\ee
The result still depends on the reduction (the pregroup version of  syntactical analysis), because it is the reduction that chooses the types $T_i$. Evidently, this definition of compositional models avoids problem of variable tensor powers.
\section{Logical functional models as functors}
The logical functional models are extensions of first order models to vector spaces. The resulting logic properly extends first order logic, because not only individuals but also sets of individuals have truth values.

A functor $\cF$ from the lexical category to the category of finite-dimensional vector spaces over the field of real numbers is a \pc{logical functional model} if  it maps the sentence type $\gs$  to a two-dimensional space $S=  \cF(\gs)$ with  `canonical' basis vectors $\top$ and $\bot$, nouns to sums of basic vectors of $N=\cF(\gn)$, determiners and attributive adjectives to  projectors, verbs and predicative adjectives to `predicates' and logical words to `logical connectives' .

\noindent{\sc{predicates}}\\
Let $A=\{a_1,\dots,a_n\}$ be an orthonormal basis of $\bbR^n=V_A$\. A linear map $p: V_A   \to S$ is a \emph{predicate} if $p(a) \in \{0, \top, \bot\}$ for any $a\in A$. It is a  \emph{predicate on $A$} if $p(a)\neq 0$ for all $a\in A$. \\
 
\noindent Examples are the linear maps $\mt{true}: V_A   \to S$ and  $\mt{false}: V_A   \to S$ satisfying 
$$\mt{true}(a_i)=\top\ \text{respectively}\  \mt{false}(a_i)=\bot\  \sn{i}\.$$ 

A linear predicate $p$ `counts' the number of basis vectors for which it takes the value $\top$. Identify any subset $B= \{a_{i_1}, \dots, a_{i_m}\}$  of $m$ distinct basis vectors  with the vector  $\oa B=\sum_{l =1}^m a_{i_l} $\,. Let $n_{pB}$ be the number of elements of $ B$ for which $p$ returns the value $\top$. Assume that $p$ is a predicate on $A$. The following holds \\
{\sc{counting property}}
\be\label{CountingProperty} p(\oa B) = n_{pB} \top +(m - n_{pB}) \bot  \ \text{and}\  n_{\mt{true}B}=|B|\.\ee

The counting property gives us a clue about how to generalise truth-values to real vector spaces.  Recall that a vector $v$  is co-linear to a vector $w$ if there is a scalar $\alpha \neq 0$ such that $v=\alpha w$\\
\noindent{\sc{truth-values}}\\
Let $p$ be a linear predicate  on $A$ and $X$ any vector of $V_A$.  We say that 
\bco 
\item $p(X)$ is \emph{true} if $p(X)$ is co-linear to $\top$  
\item $p(X)$ is \emph{false} if $p(X)$ is co-linear to  $\bot $ 
\item $p(X)$ is \emph{mixed} if $p(X)=\alpha  \top  + \beta \bot\ \text{for some }\  \alpha \neq 0, \beta \neq 0$ 
\item   $p(X)$ is \emph{mute} if  $p(X)=0$. 
\eco  
The corresponding logic has four truth values, namely `true', `false', `mixed' and `mute'.  A linear predicate assigns to a basis vector (individual) either `true' or `false' or `mute'. If the latter is the case, the predicate has no answer to the question whether the individual has the property or not.  

Truth-values are invariant under scaling. The vectors $X$ and $\lambda X$ have identical truth-values for  $\lambda \neq 0$. 
Saying `$p(X)$ is not true'   means that $p(X)$ is not co-linear to the basis vector $\top$\.   This does not imply that `$p$ is false on $X$'.\\
\noindent{\sc{logical connectives}}\\
The \emph{logical connectives} are the linear maps  $\mt{not} :S   \to S$,   $\mt{and} :S\ot S   \to S$,  $\mt{or} :S\ot S   \to S$ and  $\mt{ifthen} :S\ot S   \to S$ determined by their values on the basis vectors $z \in \{\top\ot \top,\top\ot \bot, \bot\ot \top,\bot\ot \bot \}$ thus 
$$ \ba{llll}
\mt{and} (z) &= \bc\top & \text{if}\  z =\top\ot \top\\ \bot&\text{else}\ec & \quad \mt{or} (z) &= \bc\bot  &\text{if}\   z =\bot\ot \bot\\
\top & \text{else} \ec\\
 \mt{ifthen} (z) &=\bc \bot  & \text{if}\  z=\top\ot \bot\\    \top & \text{else} \ec&\quad \mt{not} (\top) &=\bot  \quad  \mt{not} (\bot) = \top\,.\ea$$
   The logical connectives induce a Boolean algebra structure on the set of predicates on $A$ with largest element $\mt{true}$. Let $d_A: V_A \to V_A \ot V_A$ be the unique linear map satisfying  
 $d_A(a_i)= a_i \ot a_i\ \sn{i}$
 and
 $$\la p,  q \ra =(p\ot q) \circ d_A: V_A \to  S \ot  S\.$$
Then the linear maps
\bes\label{logicalConnectives}
\mt{not} \circ p :V_A  \to S, \  \mt{and} \circ \la p,q\ra, \  \mt{or}\circ \la p,q \ra,\ \mt{ifthen}\circ \la p,q \ra :V_A  \to S\ees
are predicates on $A$. 

\noindent{\sc{logical consequence relation}}\\
A predicate $q$ is said to be a \emph{logical consequence} of a predicate $p$ if 
$$\mt{ifthen}  \circ \la p,q \ra =  \mt{true}\.$$ 

The logic introduced above extends first order predicate logic. Indeed,  
assume that a vector $X=\sum_{i=1}^n \alpha_{i} a_{i} \neq 0$  satisfies $\alpha_{i}\geq 0$, $\sn{i}$, and let $\overline X = \{a_{i_1}, \dots a_{i_m}\}$ be the subset of  basis vectors $a_{i_k}$ for which $\alpha_{i_k}\neq 0$\.   Then   the following holds \\
\\
{\sc{fundamental property}} 
\be\label{fund}\ba{ll} 
 p \  \text{is true on}\  X  &\Leftrightarrow \forall x(x\in \overline X \Rightarrow p(x)= \top)\\
 p \  \text{is false on}\  X    &\Leftrightarrow \forall x(x\in\overline X \Rightarrow p(x)= \bot) 
\\
 p \  \text{is mixed on}\  X  &\Leftrightarrow \exists x \exists y (x,y\in \overline X  \  \& \  p(x)= \top\  \&  \ p(y)= \bot)\.
 \ea\ee
Words are interpreted by vectors with non-negative coordinates in the functional vector models.  Hence, the Fundamental Property applies to all of them.

The linear map $\mt{not}$ plays a double role in this logic. It is negation when the predicate is applied to a basis vector. For instance, \emph{joe is tall} versus \emph{joe is not tall}. It is the opposite, when applied to an arbitrary vector, a second order entity. For instance, \emph{all boys are tall} versus \emph{no boys are tall}. The latter assertion implies the negation of the former, but the converse does not hold. 
  
{\exa{\label{functional}{Property? : yes/no} } }\\
Consider a game involving chips that come in different shapes and colours.   Each shape is coloured with one or several of the colours \pc{red},  \pc{yellow} and  \pc{blue}.  
The machine that distributes the chips can recognise colours, but not shapes. Players who want a certain shape therefore must describe the shape, say   \pc{triangle}, \pc{square}, \pc{circle}, in terms of colour combinations.

A player who believes in functional models observes thirty chips $A=\{a_1, \dots, a_{30}\}$ extracted from the machine and represents them by a functional model  $\cF$, namely 
 \bes\ba{c}
\cF(\gn)=N=V_A\\
 \cF(\mt{triangle})  =  a_1+ \dots +a_{10}\quad \cF(\mt{square})  =  a_{11}+ \dots +a_{20}\quad \cF(\mt{circle})  =  a_{21}+ \dots +a_{30}   \\
\ba{llll}
 \cF(\mt{new})(a_i) &= a_i \ \text{if } i =5, 7-15,20,25,30 &\ \cF( \mt{new})(a_i)& = 0 \ \text{ else }\\
  \cF( \mt{blue})(a_i)& = \top \ \text{if } i=16- 20&\ \cF( \mt{blue})(a_i)& =\bot \ \text{ else }\\
\cF(\mt{red})(a_i)& = \top \ \text{if } i = 1-9, 11 - 20  & \ \cF( \mt{red})(a_i)& =\bot \ \text{ else }\\
\cF(\mt{yellow})(a_i) &= \top \ \text{if } i = 7-15,  21 &\ 
 \cF(\mt{yellow})(a_i) &= \bot \ \text{ else }.\ea
\ea\ees
Because of his preference for new chips, he computes the noun phrases \pc{new triangles}, \pc{new squares} etc. 
  $$\ba{ll}
  \cF(\mt{new}\circ\mt{square}) &= \cF(\mt{new})\circ\cF(\mt{square})
  =\cF(\mt{new}) (a_{11}+ \dots +a_{20})\\
  &=a_{11}+ a_{12}+a_{13}+a_{14} +a_{15}+a_{20}\\
    \cF(\mt{new}\circ\mt{triangle}) &=   \cF(\mt{new})\circ\cF(\mt{triangle})= \cF(\mt{new}) (a_{1}+ \dots +a_{10})\\
    &=a_{5}  +a_{7}+a_{8}+a_{9}  +a_{10}\\
    \cF(\mt{new}\circ\mt{circle}) &=\cF(\mt{new})\circ\cF(\mt{circle})=\cF(\mt{new})  (a_{21} +\dots+a_{30})\\
    &=a_{25} +a_{30}\ea$$  
Concentrating on \pc{triangles}, he finds that the meaning of the sentence `\pc{No triangles are blue}' computes to $\mt{not}\circ \mt{blue} \circ \mi{in}_{\gcp\gn} \circ \mt{triangles}$ in the lexical category, by \re{meaning}.  Hence, the interpretation of the sentence in the functional vector space model $\cF$ maps logical words to logical connectives and the inequalities to identities. Therefore $\cF(\mt{not})=\mt{not}$ and the meaning of the sentence in the model $\cF$ is
$$
 \mt{not} \circ \cF(\mt{blue})\circ \cF(\mt{triangle}) =  \mt{not}(10\cdot \bot) =10 \cdot \top.$$
 The resulting vector is colinear to $\top$, hence the sentence \pc{No triangles are blue} is true in the model. The Fundamental Property implies that $\cF(\mt{blue})(x) =\bot$ for every basis vector $x \in \cF(\mt{triangle})$.
The predicates  $\cF(\mt{red})$ and $\cF(\mt{yellow})$ are mixed on $\cF(\mt{triangle})$
$$\ba{l@{\ =\ }l} 
 \cF(\mt{red}\circ\mt{triangle})& \cF(\mt{red})( a_1+ \dots  +  a_{9}) + \cF(\mt{red})( a_{10})=    9\cdot\top+1\cdot\bot  \\
  \cF(\mt{yellow}\circ\mt{triangle})&  4\cdot\top  + 6 \cdot\bot  \ \.
 \ea $$
{\exa{\label{conceptual}{Property? : probability of yes} } }\\
The player decides to describe the concepts  \pc{triangle}, \pc{square}, \pc{circle} by their colours and use the probability that a chip with colour combination $c_j$ has shape $p$.  Therefore he needs the number $ k_{pj}$ of chips of shape $p$ appearing with a given colour combination $c_j$ and the number $m_j$ of chips having colour combination $c_j$.  Using $\pt{r}$ if the red colour is present and $\neg \pt{r}$ if the red colour is absent and similarly for the other colours, he first arranges the chips according to their colour combinations  
 \bes  
 \ba{lll}
 \text{combination}&\text{chips }&\text{number}\\
\ba{llll} 
c_1= &\  \  \pt{r} & \neg \pt{y} &\ \, \pt{b}\\
c_2= & \ \ \pt{r} &\  \ \pt{y} &   \neg\pt{b}\\
c_3= &\ \  \pt{r} &\neg\pt{y} &  \neg\pt{b}\\
c_4 = &\neg \pt{r} & \ \, \pt{y}& \neg \pt{b}\\ 
c_5= &\neg \pt{r} &  \neg  \pt{y}& \neg \pt{b}\ea 
&
\ba{l} 
C_1=\{a_{16} ,\dots,a_{20}\}\\
C_2=\{a_7,a_8,a_9\}\cup\{a_{11} ,\dots,a_{15}\}\\
C_3= \{a_{1} ,\dots, a_{6}\}\\
C_4= \{a_{10}, a_{21}\}\\
C_5= \{a_{22} ,\dots, a_{30}\}  \ea 
&
\ba{l} 
m_1= 5\\ 
m_2 =  8\\
m_3 = 6 \\
m_4=  2 \\
m_5 =  9  \.
\ea
 \ea\ees   

\bes
\ba{llll}
\text{number}&\text{shape}&\text{combination}\\
 \ba{l}
k_{s1}=5  \\
k_{s2} = 5\\
k_{c4} = 1 \\ 
k_{c5} = 9 \\  
k_{t3} =6  \\ 
k_{t2}  =3 \\ 
k_{t4} = 1 \ea
 &
 \ba{l} 
s=\text{square} \\
s=\text{square} \\
c=\text{circle}\\
 c=\text{circle}\\
t=\text{triangle}\\
t=\text{triangle}\\
t=\text{triangle}   \ea
 &
 \ba{llll} 
c_1= &\  \  \pt{r} & \neg \pt{y} &\ \, \pt{b}\\
c_2= & \ \ \pt{r} &\  \ \pt{y} &   \neg\pt{b}\\
c_4 = &\neg \pt{r} & \ \, \pt{y}& \neg \pt{b}\\ 
c_5= &\neg \pt{r} &  \neg  \pt{y}& \neg \pt{b}\\
c_3= &\ \  \pt{r} &\neg\pt{y} &  \neg\pt{b}\\
c_2= & \ \ \pt{r} &\  \ \pt{y} &   \neg\pt{b}\\
c_4 = &\neg \pt{r} & \ \, \pt{y}& \neg \pt{b} \ea 
\ea
\ees

Representing each shape $p$ by the extent to which the primitive properties are true, namely the vector 
$$p=\sum_j  k_{pj}/m_j \cdot c_j$$   
 he obtains 
$$\ba{lll}  
\cF(\mt{square})&\mapsto \mi {square} &= 5/5 \cdot c_1 +  5/8\cdot c_2 \\
\cF(\mt{triangle})&\mapsto  \mi {triangle} &= 3/8 \cdot c_2+6/6 \cdot c_3 + 1/2 \cdot c_4 \\
\cF(\mt{circle})&\mapsto \mi {circle} &=1/2\cdot c_4+ 9/9 \cdot c_5\,.
\ea$$
It suffices to ask for a red chip that is not yellow and not blue to obtain a triangle.
\section{Conceptual vector semantics}\label{CVSL}
 Distributional vector space models represent  words by vectors in a finite dimensional space $V$.  Its basis vectors are assimilated with previously chosen words, e.g. the key words of a thesaurus or the most frequent words in a set of documents. If the distribution is based on context, the coordinates of a word vector stand for frequencies of co-occurrences with the respective basis vectors in contexts. 
 
 This study takes a more general approach.  We assume that the coefficients of the word vectors belong to the real interval $[0,1]$. Call \emph{concept vector} any vector that has its  coordinates in $[0,1]$. The method by which the coordinates have been obtained is irrelevant in this section.  We want to define a logic on the space $V$ so that we can reason with vectors in a vector space model according to our intuition. For instance,
the vectors assigned to the statements   \pc{All apples are juicy} and \pc{No apples are juicy} must be contradictory.

The following definitions refer to an orthonormal basis $A=\{a_1, \dots, a_n\}$ of $\bbR^n$. A vector is \pc{Boolean} if its coordinates with respect to $A$ are equal to $0$ or $1$.  The set of  {Boolean vectors} is denoted 
$$\cB_A=\{\sum_{i=1}^n \alpha_i a_i\,:\, \alpha_i \in\{0,1\}\}\.$$
 Denote $\name f_{AA}$ the matrix in the basis $A$ defined by an endomorphism $f$. The endomorphisms of $\bbR^n$ that are diagonalisable with respect to $A$ and the projectors among them form the sets
$$\ba{ll}\cD_A &=\{f:\bbR^n \lra \bbR^n\,:\, \name f_{AA} \ \text{is a diagonal matrix}\}\\ \cP_A &= \{ f\in \cD_A\,:\, f\circ f=f\} 
\.\ea$$ 
There is an obvious bijection between vectors of $\bbR^n$ and $\cD_A$ via the correspondence    
$$ X=\sum_{i=1}^n \alpha_i a_i \leftrightarrows D_X = \begin{pmatrix}
\alpha_{1} &  & 0\\
  &\ddots &   \\
0  & &\alpha_{n}\\
\end{pmatrix} $$ 
This  correspondence is the vector space analogue of the bijection between subsets and predicates in set theory.  Moreover, it maps $\cB_A$ onto $\cP_A$.  \\

The algebraic connectives below generalise the connectives of predicate logic and coincide with those of quantum logic for projectors arising in vector space models.  They are inspired by a probabilistic interpretation of truth. \\
\noindent{\sc{algebraic connectives}}\\
 The  \pc{algebraic connectives} are defined for scalars and for arbitrary square matrices thus
\bes \label{algebraicConnScal}
\ba{ll@{\ =\ }l@{\qquad }l@{\ =\ }l}
\text{negation}&\neg \alpha&1-\alpha &\neg D&1-D\\
\text{conjunction}&\alpha \,{\land}\, \beta &\alpha \beta &D \,{\land}\, E &D E\\
\text{disjunction}&\alpha \,\lor\,  \beta&\alpha + \beta  -\alpha  \beta  &D \,\lor\,  E&D + E  - D   E\\
\text{implication}&\alpha \rightarrow  \beta  & 1-\alpha +\alpha  \beta  &D \rightarrow  E  & 1-D +D   E \quad .\ea\ees 
The algebraic connectives are lifted from scalars to vectors by the conditions
 \bes\label{VectorConnective} \ba{l@{\ =\ }l@{\ =\ }l}
\neg X&\neg(\sum_{i=1}^n  \alpha_i   a_i )  & \sum_{i=1}^n (\neg\alpha_i  ) a_i  \\  
X\triangledown Y&(\sum_{i=1}^n  \alpha_i   a_i ) \triangledown  (\sum_{i=1}^n \beta_i a_i ) & \sum_{i=1}^n (\alpha_i \triangledown \beta_i) a_i\,,\ea\ees
where  $\triangledown$ stands for any of the binary algebraic connectives. It follows that 
\be \label{MatrixConnective}  \neg D_X  =D_{\neg X }  \quad \text{and} \quad D_X \triangledown D_Y =D_{X \triangledown Y}\ee 
The  two equalities above say that the one-to-one correspondence that identifies a vector $X$ with the diagonal matrix $D_X$ is an isomorphism.
Note that the conjunction of two vectors is the same as their pointwise product and that they relate to composition by the equalities
\be\label{Compo} D_{X} \circ D_Y= D_{X }\land D_{Y}= D_{X \land Y}= D_{X \odot Y},  \qquad D_{X} \circ \bra Y=\bra {X \land Y}\,,
\ee 
where  $\bra{Z}$ denotes the matrix of the linear map from $I$ to $V_A$ that assigns  the vector $Z\in V_A$ to the basis vector of $I$. 

Concept vectors are closed under the algebraic connectives, because the interval $[0,1]$ is closed under the algebraic connectives on scalars. 
The proof is straightforward.  For instance, to show the assertion for the algebraic disjunction,
assume $\alpha, \beta \in [0,1]$.  The inequality $0 \leq \alpha+\beta  -\alpha  \beta $ follows from $\beta -\alpha  \beta =\beta(1 -\alpha)\geq 0$.  The inequality $ \alpha+\beta  -\alpha  \beta \leq 1$ follows from $1-\alpha -\beta +\alpha  \beta =(1-\alpha)(1-\beta)\geq 0$\.

The algebraic connectives do not define a lattice structure,  the algebraic conjunction for instance is not idempotent unless the involved scalars are equal to $0$ or $1$.  The algebraic connectives  have, however,  several properties with a logical flavour, among them the laws of a weak conditional logic in the sense of   \cite{ri}. 
 
Any real numbers $\alpha, \beta \in [0,1]$ and diagonal matrices $D,E  \in \cD_A$ with entries in $[0,1]$ satisfy
\be\label{consequence}
\alpha \rightarrow \beta= 1 \iff \alpha =0 \ \text{or}\  \beta =1
\ee
$$\ba{c}
\alpha \rightarrow \beta= 1\iff \alpha  \beta=\alpha \\
\text{If}\    \alpha \rightarrow \beta= 1  \ \text{then}\   \alpha \leq \beta\\
 D \rightarrow E= 1 \iff E \circ D=D\\
 \text{If}\    D \rightarrow E= 1  \ \text{then}\   D \leq E\.
\ea$$
\\
\noindent{\sc{algebraic consequence relation}}\\
The endomorphism defined by $E$ is  an \emph{algebraic consequence} of that defined by $D$ if and only if
$$D \rightarrow E=1\.$$ 
\noindent{\sc{probabilistic consequence relation}}\\The endomorphism defined by $E$ is  a  \emph{probabilistic  consequence}  of that defined by $D$ if  $$D \leq E\.$$  

Projectors stand for properties in quantum logic and geometrical operations define the connectives.  These geometrical operations are introduced in \cite{ri}  via the range of the involved projectors  based on the fact that for every subspace there is a unique projector which maps the whole space onto the subspace in question.   
 \\ 
\noindent{\sc{geometrical connectives}}\\
Let $p,q:\bbR^n \lra \bbR^n$ be projectors.  Then  

the \emph{geometrical negation} $ \neg p$ is the unique projector that has range 
$$(\neg p)(\bbR^n)=p(\bbR^n)^\bot$$

the \emph{geometrical conjunction}  $  p\land q$ is the unique projector that has range 
$$( p\land q)(\bbR^n) = p(\bbR^n) \cap q( \bbR^n) $$

the \emph{geometrical disjunction} $  p\lor q$ is the unique projector that has range 
$$( p\lor q)(\bbR^n) = p(\bbR^n) + q( \bbR^n) $$

the \emph{geometrical implication}  $p\Rightarrow  q$ is the unique projector that has range
$$(p\Rightarrow  q)(\bbR^n) =\{ x \in \bbR^n\,:\, (q\circ p)(x) =p(x)\} \.$$
\noindent{\sc{quantum consequence relation}}\\
Projector $q $ is said to be a \emph{geometrical consequence} of projector $ p$  if and only if 
$$p\Rightarrow q =1\.$$
 
The definition makes the detour via the subspaces, because there is no obvious algebraic operation defining the projector.  For example, $p\circ q$ maps  $\bbR^n$ 
onto the intersection of the image of $p$ and the image of $q$, but $p\circ q$ is not a projector unless $p$ and $q$ commute. If $p$ and $q$ do not commute, there is no  basis in which they are both diagonalisable.  Are we not losing representatives of properties in probability when replacing the projectors by  $\cD_A$? The answers is that to the contrary, we are gaining representatives at least as long as we accept the geometric consequence relation.
\begin{prop}
If projector $q $ is a geometrical consequence of projector $ p$ then there is  an orthogonal basis of $\bbR^n$ consisting of eigenvectors of both $p$ and $q$\. 

The geometrical consequence relation and  the algebraic consequence relation coincide on projectors. If one of $p$ or $q$ is a geometrical consequence of the other then the geometrical connectives coincide with the algebraic connectives for $p$ and $q$.
\end{prop}
\pro{(Outline) Clearly, the second statement follows from the first.  To see the first statement, assume that  $q$ is a geometrical consequence of $p$  and  let $A$ be an orthonormal basis of $\bbR^n$ formed by eigenvectors of $p$\.  The  eigenvectors in $A$ left invariant by $p$ are also left invariant by $q$. Let $A_1$ denote this set and $V$ be the subspace generated by $A_1$. Then $q$ maps the orthogonal complement of $V$ onto itself whereas $p$ maps it to $0$. Hence any set $C$ of orthonormal eigenvectors of $q$ belonging to $V$ are also eigenvectors of $p$.  Thus $A_1\cup C$ is a basis of orthonormal eigenvectors for both $p$ and $q$.
}
The proposition above also implies that the algebraic connectives can be captured by geometrical properties, at least on Boolean vectors. In particular, two distinct basis vectors contradict each other. This raises the question of how to choose the basis vectors so that they represent contradictory properties.

\section{Distributional interpretations}\label{DI}
Everyday language switches commonly from asserting facts about some real or possible world to updating the concepts intervening in the statements about the facts.  This switch is related to the canonical distribution based on the counting property of  predicates. The coefficient of a property, say \pc{apple}, at a basis vector, say \pc{juicy}, is the probability of the event \pc{apple} given the event \pc{juicy}. It gives us the extent to which the property \pc{juicy} is characteristic  for the concept \pc{apple}.

The section concludes with sufficient conditions for concept logic to be reflected and predicate logic to be  preserved. 

Predicates $q_1, \dots, q_k$  on  $A=\{a_1, \dots, a_n\}$  are said to  \emph{partition $A$} if for any $a \in A$ there is a  $ j$ for which  $q_j(a) =\top$ and $i \neq l$ implies $q_i(a) =\bot$ or $ q_l(a) =\bot$  for ${i},\ l=1, \dots, k$ .    
Clearly, families of partitioning predicates are in one-to-one correspondence with set-theoretical partitions $\{C_1,\dots, C_k\}$ of  $A$ given by
$$C_j=\{a \in A\,:\, q_j(a)=\top\},\  j=1,\dots,k \.$$ 
Let $m_j=|C_j|$ so that $\sum_j m_j=n=|A|$\.  Assuming that every individual  in $A$ has probability $1/n$, the real number
 $$\mu_j =m_j/n $$
can be understood as the probability that an arbitrary individual $a \in A$ has property $q_j$,  $\sn[k]{j}$. 

Choose some orthonormal basis $C=\{c_1,\dots, c_k\}$ of $V_C=\bbR^k$. Think of the basis vectors   $c_1,\dots, c_k$ as `basic events on $A$' or as `basic properties' of the elements of $A$.  The density operator defined by the diagonal matrix 
$$D_\mu=\begin{pmatrix}\mu_1&  & 0\\
  &\ddots &   \\
0  & &\mu_{k}\end{pmatrix}$$ 
summarises the first order model consisting of $A$ and the predicates $q_j$, for $ j=1, \dots, k$\. 
The  composite of the density operator $D_\mu$ with the  $j$'th  projector of $\bbR^k$ maps an arbitrary vector $\sum_{i=1}^k\beta_i c_i$ to $\mu_j\beta_j$. We shall define for any property $p$ on $A$  a vector $\cJ(p)=\sum_{i=1}^k\alpha_i c_i\in V_C$ such that $\sum_{j=1}^k\alpha_i\mu_i =n_{pA}/n$. Recall that $n_{pA}$ is the number of elements of $A$ satisfying $p$.  If we call  $ p$ a `state' of the `system' $A, \, q_1,\dots,  q_k$ then $n_{pA}/n$ is the probability that the system is in state $p$. 

For any  predicate $p$ on $A$, the integer $n_{pC_j}$ is the coefficient of   $p(\oa{C_j})$ at $\top$. Define 
\bes\label{prediag}\cJ_C(p)=\sum_{j=1}^k  \alpha_{pj} c_j, \ \text{where} \   \alpha_{p j}= \bc  n_{pC_j}/m_j &\text{if}\ m_j\neq 0\\
0&\text{else}\ec,\sn[k]{j}\.\ees
The number $\alpha_{pj}$ is the conditional probability that an  element has property $p$ given  property $q_j$. 
It follows from the linearity of $p$ that $n_{pA}=\sum_j n_{pC_j}$.
Therefore, the probability that an arbitrary element of $A$ has property $p$  is equal to  
$$n_{pA}/n  =\sum_j n_{pC_j}/ n   =\sum_j (n_{pC_j}/m_j)(m_j/n) = \pc{trace}(D_\mu\circ D_{\cJ(p)})=\sum_j \alpha_{pj}\mu_j\.$$ 

The interpretation $\cJ_C$ is a one-to-one map from predicates on $A$ to Boolean vectors of $V_C\simeq V_A$ if $C$ consists of the singleton sets $\s{a_1}, \dots, \s{a_n}$. If this is the case then  $\alpha_{pj}=1$ if $p(a_j)=\top$ and $\alpha_{pj}=0$ otherwise. In the general case, $\cJ_C$ is neither one-to-one nor onto the set of vectors with coefficients in $[0,1]$. The following Lemma describes the general situation.
\begin{lem}\label{sufficient}
Let $p, q:V_A \to S$  be any predicates on $A$, assume that the sets $C_j$ are not empty $\sn[k]{j}$  and that for every $j$  at least one of $p$ or $q$ is constant on $C_j$. Then the following holds  
\bes \label{preservebinop} \ba{r@{\ = \ }l}
\cJ_{C}(\mt{and} \circ \la p , q\ra )  & \cJ_C( p) \land \cJ_C( q) \\
\cJ_{C}(\mt{or} \circ \la p , q\ra)   & \cJ_C( p) \lor \cJ_C( q)\\
\cJ_{C}(\mt{ifthen} \circ \la p , q\ra)  & \cJ_C( p) \rightarrow \cJ_C( q) \.\ea\ees
\end{lem}
\begin{theo}\label{constant}
Suppose that the the non-empty sets $C_1, \dots, C_k$ partition $A$ and that  $p$ and $q$ are predicates on $A$. Then  $\cJ_C$ preserves negation
\bes\label{negation}
\cJ_C(\mt{not}\circ p)=\neg\cJ_C(p)
\ees
and reflects the consequence relation 
\be\label{reflecttruth}   \cJ_C(p) \to \cJ_C(q)=1\ \text{implies}\ \mt{ifthen} \circ \la p , q\ra=\mt{true} \ee

If one of  $\cJ_C(q)$ and $\cJ_C(p)$ is an algebraic/geometrical consequence of the other then $\cJ_C$ preserves the logical connectives, i.e. the algebraic/geometrical connectives preserve the probabilistic interpretation of concept vectors.
\be \label{binop} \ba{r@{\ = \ }l}
\cJ_C( p) \land \cJ_C( q) & \cJ_{C}(\mt{and} \circ \la p , q\ra )  \\
\cJ_C( p) \lor \cJ_C( q)   & \cJ_{C}(\mt{or} \circ \la p , q\ra) \\
 \cJ_C( p) \rightarrow \cJ_C( q) & \cJ_{C}(\mt{ifthen} \circ \la p , q\ra)  \\
\cJ_C( q) \rightarrow \cJ_C( p) & \cJ_{C}(\mt{ifthen} \circ \la q , p\ra)  \.\ea\ee
\end{theo}
\pro{ 
Assume that $\cJ_C(p) \to \cJ_C(q)=1$\.  Then $1-\alpha_{p j} +\alpha_{p i}\alpha_{q j }=1$ and therefore $\alpha_{p j}=0$ or $\alpha_{q j }=1$, $\sn[k]{j}$\. Otherwise said, $p$ maps every element of $C_j$ to  $\bot$ or $q$ maps every element of $C_j$ to $\top$. As every element of $ A$ belongs to some $C_j$, we have
$$\mt{ifthen} \circ \la p , q\ra(x)= \top, \  \text{for all}\  x \in A\.$$
The  equality $\mt{ifthen} \circ \la p , q\ra=\mt{true} $ follows. This completes the proof of \re{reflecttruth}.

The equalities \re{binop} hold, because the assumptions of the preceding lemma are satisfied. 
}
Negation is preserved exactly when $D_\mu$ is positive definite, i.e $\mu_j>0$, for $j=1,\dots,k$.  This condition alone is not sufficient for $\cJ_C$ to preserve the binary connectives. 

In the particular case where the partition  of $A$ is $C_1=\{a_1\}, \dots, C_n=\{a_n\}$, the hypotheses of Lemma \ref{sufficient} and Theorem \ref{constant} are satisfied. Therefore,  $\cJ_A$ is an isomorphism of the Boolean algebra of predicates on $A$ onto the Boolean algebra of Boolean vectors $\cB_A$ of $V_A$.   
Hence in the case where $\mu_i=1/n, \sn{i}$, predicate logic, quantum logic and vector space logic are the same, because the lattice of predicates on $A$, the lattice of projectors in $\cP_A$ and the lattice of Boolean vectors $\cB_A$ are isomorphic. Composition of projectors, conjunction of predicates and pointwise product of vectors are three variants of the same operation on property words.

In the general case, extend $\cJ_C$ from predicates on $A$ to projectors and Boolean vectors of the space $V_A$.  Then Theorem \ref{constant} remains valid. Moreover, the composite  $\cJ_C\circ \cF$ is a well defined map from property words  to  vectors in $V_C$. The induced compact closed structure preserving functor $\cM_C:\cL(\cB)\to V_C$ given by
$$\cM_C(\mt{word}_T) =\cJ_C(\cF(\mt{word}_T))$$
maps the lexical meaning of a string to the pointwise product of the word vectors, by Equation \re{pointwise}.  
The truth-probabilistic relation between the vector model and the logical functional model extends under the sufficient conditions of Theorem \ref{constant}  to string of words  
$$\cM_C( \mt{word}_{1T_1}) \odot \dots \odot\cM_C( \mt{word}_{nT_n})   =\cJ_C(\cF(r\circ(\ol{word}_{1T_1}  \ot \dots \ot\ol{word}_{nT_n} ))).$$

Theorem 1 also provides a method for checking  how appropriate a possible extension of an arbitrary vector model to other noise words would be. For example, $\cF(\mt{are})$ is the identity map in all functional  models.  Hence  $\cF(\emph{rocks are grey})= \cF(\mt{grey})\circ \cF(\mt{rocks})$. The unit for $\odot$, the vector $\oa1$, is the identity when we think of $\odot$ as composition. Hence we let $\cM(\mt{are})=\oa1$ so that in the vector model 
$$\cM(\mt{rocks})\odot \cM(\mt{are})\odot \cM(\mt{grey})=\cM(\mt{rocks} )\odot \cM(\mt{grey})=\cM(\mt{rocks} )\land \cM(\mt{grey})\.$$   
Similarly, if $\cM(\mt{and})=\oa1$ then $\cM$ maps the lexical meaning $\mt{and}\circ(\mt{red} \ot\mt{blue}):\gp\ot\gp \to \gp$ of the string \emph{red and blue} to 
$$\cM(\mt{and})\odot\cM(\mt{red}) \odot\cM(\mt{blue})= \cM(\mt{red}) \odot\cM(\mt{blue}) =\cM(\mt{red}) \land\cM(\mt{blue})\.$$

Our toy example concerns unary predicates only, for the sake of simplicity, but the definitions apply to arbitrary predicates.  Consider the case of binary predicates.
A logical functional model $\cF$ interprets a transitive verb as a binary predicate $p:V_{E\otimes E }\simeq V_E\ot V_E \to S$ and nouns as vectors in $V_E$.  A somewhat realistic pregroup lexicon lists a transitive verb with type $\gn_{\mt{sub}}^r\gs\gn_{\mt{ob}}^\ell$ and a noun with the types $\gn_{\mt{sub}}$ and $\gn_{\mt{ob}}$. Note that the tensor product of vectors $v,w \in V_A$ is related to the pointwise product in $V_{E\otimes E}$ by the equality
$$v\ot w =(v\ot \oa1)\odot(\oa1 \ot w )$$
The passage from $\cF$ to the vector space model $\cM$ described above is facilitated  by defining 
$$ M( \mi{noun}:{\gn_{\mt{sub}}}) =v \ot \oa1\qquad M( \mi{noun}:{\gn_{\mt{ob}}})=\oa1 \ot w \,,$$ 
where $v=\cJ(\cF(\mt{noun}_{\gn_{\mt{sub}}}))$ and $w =\cJ(\cF(\mt{noun}_{\gn_{\mt{ob}}}))$.   

The fact that the binary operation$\odot$ is  commutative  does not imply that meanings are necessarily commutative, because the model interprets the sentence \emph{cats chase dogs} by
$$ \cM( \mt{cats}_{\gn_{\mt{sub}}})\odot \cM(\mt{chase}_{\gn_{\mt{sub}}\gs\gn_{\mt{ob}}^\ell})\odot\cM(\mt{dogs}_{\gn_{\mt{ob}}})\,,$$
and \emph{dogs chase cats} by  
$$\cM( \mt{dogs}_{\gn_{\mt{sub}}})\odot \cM(\mt{chase}_{\gn_{\mt{sub}}^r\gs\gn_{\mt{ob}}^\ell})\odot\cM(\mt{cats}_{\gn_{\mt{ob}}})\.$$  
\section{Logic and the basis of the vector space}
Theorem 1 in the preceding section shows that the quality of reasoning  in a vector space model depends essentially on the choice of the basic concepts.
The most frequent property words in a document do not in general constitute a partition.  In the world of fruit, there may be things that are \emph{juicy} and \emph{sweet} simultaneously. It is the requirement borrowed from quantum logic that orthogonal vectors must be contradictory that forces basis vectors to be contradictory.  

There is however a general method for transforming an arbitrary choice of properties into a set of partitioning properties.  This method was used in our  Example \ref{conceptual}. 
 
  Let $P= \s{\se[d]{{w}}}$ be a set of property words in the lexicon.  Think of them as primitive properties. Invent a two-dimensional space $S_ i=V_{\{{\pt w}_ i, \neg {\pt w}_ i\}}$ with basis vectors ${\pt w}_ i$ and $\neg {\pt w}_ i$, $\sn[d]{i}$ and define %
the \emph{concept space generated by} $P$ as 
$$C(P) = S_ 1\ot \dots \ot S_ d\.$$
The basis vectors of $C(P)$ are of the form   $$c_j= c_j(1) \ot \dots \ot c_j(d) $$
where  $c_j(i)\in \s{\mt{w}_ i, \neg  \mt{ w}_ i}, \sn[2^d]{j},\  i=1,\dots,d$.

Without loss of generality, we may assume that the functional model interprets the  words  $\se[d]{{w}}$ as predicates  $\se[d]{p}$ on a space $V_A$.  The basis vectors correspond to the following partition $\se[2^d]{C}$ of subsets of $A$ 
$$a \in C_j \iff p_i(a)=\bc 
\top& \text{ if } c_j(i) = \pt w_i  \\ 
 \bot & \text{ if }\neg \pt w_i\ec, \sn[d]{i}\.$$ 

If we work within the subspace $C$ of $C(P)$ generated by the basis vectors $c_j$ for which $C_j \neq \emptyset$, the interpretation $\cJ_{C}$ reflects concept logic. It preserves predicate logic under the conditions of Theorem \ref{constant}.  

{\exa{\label{observation}\emph{} } }\\
A text describing the world of coloured chips may contain pertinent knowledge about the relation of shapes and colours. The concept triangle to be characterised by the three primitive properties \emph{red},   \emph{yellow} and  \emph{blue}   has the form
 $$ \mi{triangle} =  \alpha_{1}c_1+\cdots + \alpha_{8} c_8,\, 0 \leq \alpha_{i} \leq 1  \. $$  
 The truth of the statement \pc{No triangle is blue} implies that $\alpha_i=0$  for all  four colour combinations $c_i$ involving  \pt{b} (blue) without the negation symbol. Hence
  $$ \mi{triangle} =  \alpha_{2}c_2+\alpha_{3} c_3 +  \alpha_{4} c_4 +  \alpha_{5} c_5 \,. $$
Note that the vector $ \mi{triangle}$ is orthogonal to every basis vector that lists the colour blue as present, namely $c_1,c_6,c_7$ and $c_8$\.  Therefore $ \mi{triangle}$ is orthogonal to the subspace `blue' generated by $c_1,c_6,c_7$ and $c_8$.  

The player wants to find out if there is also a winning strategy for a new square. He computes the concept corresponding to the predicate $\cF(\mt{new}) $ and then the concept $\mi{new} \,\mi{square}$
$$\ba{c}\cJ(\cF(\mt{new}) ) = 1/5\cdot c_1+ 8/8\cdot c_2+1/6\cdot c_3 + 1/2 \cdot c_4+2/9\cdot c_5\\
\cJ(\cF(\mt{new} \circ \mt{square}))   = 1/5\cdot c_1 +5/8\cdot c_2= \cJ(\cF(\mt{new})) \odot \cJ(\cF( \mt{square}))\.\ea$$
He guesses that the same works for the other shapes.   But
$$ \cJ(\cF(\mt{new} \circ \mt{triangle}))= 3/8\cdot c_2+1/6\cdot c_3 +1/2\cdot c_4 \neq\cJ(\cF(\mt{new})) \odot \cJ(\cF(\mt{triangle})) .$$ 
In our particular characterisation of the world of chips the concept `new' does not always interact logically with the other concepts.

\section{Conclusion}
The preceding is only an outline how to extend vector space models compositionally to statements that go beyond the property words.  The vector space models depend on the chosen probability on the vector space standing for the `universe of discourse'. Grammar is another parameter of our vector space models. Different types may result in different meanings. The parameter `background knowledge' is also present via the choice of the primitive properties leading to the basis vectors of the concept space.  

The truth-probabilistic approach to compositional vector space models provides a tool to compare probabilistic reasoning in vector space models and reasoning with traditional logical tools, depending on the parameters. 

\nocite{*}
\bibliographystyle{eptcs}

\begin{thebibliography}{10}
\providecommand{\bibitemdeclare}[2]{}
\providecommand{\surnamestart}{}
\providecommand{\surnameend}{}
\providecommand{\urlprefix}{Available at }
\providecommand{\url}[1]{\texttt{#1}}
\providecommand{\href}[2]{\texttt{#2}}
\providecommand{\urlalt}[2]{\href{#1}{#2}}
\providecommand{\doi}[1]{doi:\urlalt{http://dx.doi.org/#1}{#1}}
\providecommand{\bibinfo}[2]{#2}

\bibitemdeclare{book}{as11}
\bibitem{as11}
\bibinfo{author}{Nicholas \surnamestart Asher\surnameend}
  (\bibinfo{year}{2011}): \emph{\bibinfo{title}{Lexical Meaning in Context}}.
\newblock \bibinfo{publisher}{Cambridge University Press}, \doi{10.1017/cbo9780511793936}.

\bibitemdeclare{inproceedings}{ccs}
\bibitem{ccs}
\bibinfo{author}{Stephen \surnamestart Clark\surnameend}, \bibinfo{author}{Bob
  \surnamestart Coecke\surnameend} \& \bibinfo{author}{Mehrnoosh \surnamestart
  Sadrzadeh\surnameend} (\bibinfo{year}{2008}): \emph{\bibinfo{title}{A
  Compositional Distributional Model of Meaning}}.
\newblock In \bibinfo{editor}{W.~Lawless \surnamestart P.~Bruza\surnameend} \&
  \bibinfo{editor}{J.~\surnamestart van Rijsbergen\surnameend}, editors: {\sl
  \bibinfo{booktitle}{Proceedings of Conference on Quantum Interactions}},
  \bibinfo{organization}{University of Oxford}, \bibinfo{publisher}{College
  Publications}.

\bibitemdeclare{inproceedings}{gs11}
\bibitem{gs11}
\bibinfo{author}{Edward \surnamestart Grefenstette\surnameend} \&
  \bibinfo{author}{Mehrnoosh \surnamestart Sadrzadeh\surnameend}
  (\bibinfo{year}{2011}): \emph{\bibinfo{title}{A Compositional Distributional
  Semantics, Two Concrete Constructions, and some Experimental Evaluations}}.
\newblock In: {\sl \bibinfo{booktitle}{Lecture Notes in Computer Science}},
  \bibinfo{publisher}{Springer}.
\newblock \bibinfo{note}{Pending publication}, \doi{10.1007/978-3-642-24971-6-5}.

\bibitemdeclare{inbook}{ks}
\bibitem{ks}
\bibinfo{author}{Dimitri \surnamestart Kartsaklis\surnameend},
  \bibinfo{author}{Mehrnoosh \surnamestart Sadrzadeh\surnameend},
  \bibinfo{author}{Stephen \surnamestart Pulman\surnameend} \&
  \bibinfo{author}{Bob \surnamestart Coecke\surnameend} (\bibinfo{year}{2013}):
  \emph{\bibinfo{title}{Reasoning about Meaning in Natural Language with
  Compact Closed Categories and Frobenius Algebras}}.
\newblock \bibinfo{publisher}{Cambridge University Press}.

\bibitemdeclare{article}{kl}
\bibitem{kl}
\bibinfo{author}{G.M. \surnamestart Kelly\surnameend} \& \bibinfo{author}{M.L.
  \surnamestart Laplaza\surnameend} (\bibinfo{year}{1980}):
  \emph{\bibinfo{title}{Coherence for compact closed categories}}.
\newblock {\sl \bibinfo{journal}{Journal of Pure and Applied Algebra}}
  \bibinfo{volume}{19}, pp. \bibinfo{pages}{193--213}, \doi{10.1016/0022-4049(80)90101-2}.

\bibitemdeclare{article}{kr1}
\bibitem{kr1}
\bibinfo{author}{Marcus \surnamestart Kracht\surnameend}
  (\bibinfo{year}{2007}): \emph{\bibinfo{title}{Compositionality: The Very
  Idea}}.
\newblock {\sl \bibinfo{journal}{Research in Language and Computation}}
  \bibinfo{volume}{5}, pp. \bibinfo{pages}{287--308}, \doi{10.1007/s11168-007-9031-5}.

\bibitemdeclare{inproceedings}{la99}
\bibitem{la99}
\bibinfo{author}{Joachim \surnamestart Lambek\surnameend}
  (\bibinfo{year}{1999}): \emph{\bibinfo{title}{Type Grammar revisited}}.
\newblock In \bibinfo{editor}{Alain \surnamestart Lecomte\surnameend}, editor:
  {\sl \bibinfo{booktitle}{Logical Aspects of Computational Linguistics}}, {\sl
  \bibinfo{series}{LNAI}} \bibinfo{volume}{1582},
  \bibinfo{publisher}{Springer}, \bibinfo{address}{Heidelberg}, pp.
  \bibinfo{pages}{1--27}, \doi{10.1007/3-540-48975-4-1}.

\bibitemdeclare{book}{la08}
\bibitem{la08}
\bibinfo{author}{Joachim \surnamestart Lambek\surnameend}
  (\bibinfo{year}{2008}): \emph{\bibinfo{title}{From word to sentence}}.
\newblock \bibinfo{publisher}{Polimetrica}, \bibinfo{address}{Milano, Italia}.

\bibitemdeclare{book}{mcl}
\bibitem{mcl}
\bibinfo{author}{Saunders \surnamestart Mac~Lane\surnameend}
  (\bibinfo{year}{1971}): \emph{\bibinfo{title}{Categories for the Working
  Mathematician}}.
\newblock \bibinfo{publisher}{Springer}, \doi{10.1007/978-1-4612-9839-7}.

\bibitemdeclare{inproceedings}{ml}
\bibitem{ml}
\bibinfo{author}{Jeff \surnamestart Mitchell\surnameend} \&
  \bibinfo{author}{Mirella \surnamestart Lapata\surnameend}
  (\bibinfo{year}{2008}): \emph{\bibinfo{title}{Vector-based Models of Semantic
  Composition}}.
\newblock In: {\sl \bibinfo{booktitle}{Proceedings of the 46th Annual Meeting
  of Computational Linguistics}}, pp. \bibinfo{pages}{236--244}.

\bibitemdeclare{article}{pl}
\bibitem{pl}
\bibinfo{author}{Anne \surnamestart Preller\surnameend} \&
  \bibinfo{author}{Joachim \surnamestart Lambek\surnameend}
  (\bibinfo{year}{2007}): \emph{\bibinfo{title}{Free compact
  {\textit{2}}-categories}}.
\newblock {\sl \bibinfo{journal}{Mathematical Structures for Computer
  Sciences}} \bibinfo{volume}{17}(\bibinfo{number}{1}), pp.
  \bibinfo{pages}{1--32}, \doi{10.1017/S0960129506005901}.

\bibitemdeclare{article}{ps11b}
\bibitem{ps11b}
\bibinfo{author}{Anne \surnamestart Preller\surnameend} \&
  \bibinfo{author}{Mehrnoosh \surnamestart Sadrzadeh\surnameend}
  (\bibinfo{year}{2011}): \emph{\bibinfo{title}{Semantic Vector Models and
  Functional Models for Pregroup Grammars}}.
\newblock {\sl \bibinfo{journal}{Journal of Logic, Language and Information}}
  \bibinfo{volume}{20}(\bibinfo{number}{4}), pp. \bibinfo{pages}{419--423},
  \doi{10.1007/s10849-011-9132-2}.

\bibitemdeclare{book}{ri}
\bibitem{ri}
\bibinfo{author}{C.J.~van \surnamestart Rijsbergen\surnameend}
  (\bibinfo{year}{2004}): \emph{\bibinfo{title}{The Geometry of Information
  Retrieval}}.
\newblock \bibinfo{publisher}{Cambridge University Press}, \doi{10.1017/cbo9780511543333}.

\bibitemdeclare{book}{wi04}
\bibitem{wi04}
\bibinfo{author}{Dominic \surnamestart Widdows\surnameend}
  (\bibinfo{year}{2004}): \emph{\bibinfo{title}{Geometry and Meaning}}.
\newblock {\sl \bibinfo{series}{CSLI lecture notes}} \bibinfo{volume}{172},
  \bibinfo{publisher}{CSLI Publications}.

\bibitemdeclare{inproceedings}{wi08}
\bibitem{wi08}
\bibinfo{author}{Dominic \surnamestart Widdows\surnameend}
  (\bibinfo{year}{2008}): \emph{\bibinfo{title}{Orthogonal negation in
  vector-spaces for modelling word-meanings and document retrieval}}.
\newblock In: {\sl \bibinfo{booktitle}{Proceedings of the 41st Annual Meeting
  of the Association for Computational Linguistics}}, \doi{10.3115/1075096.1075114}.
\end{thebibliography}

\end{document}